\documentclass[%
 reprint,showpacs,showkeys,nofootinbib,amsmath,amssymb, aps,]{revtex4-1}

\usepackage{graphicx}% Include figure files
\usepackage{dcolumn}% Align table columns on decimal point
\usepackage{bm}% bold math

\usepackage[dvips]{color}

\begin{document}

\title{Is dark matter made of mirror matter? Evidence from cosmological data.}

\author{Paolo Ciarcelluti}
 \email{paolo.ciarcelluti@gmail.com}
 \affiliation{Web Institute of Physics, www.wiph.org}

\author{Quentin Wallemacq}
 \email{quentin.wallemacq@ulg.ac.be}
 \affiliation{IFPA, D\'ep. AGO, Universit\'e de Li\`ege, 4000 Li\`ege, Belgium}

\date{\today}% It is always \today, today,

\begin{abstract}
We present new fast numerical simulations of cosmic microwave background and large scale structure in the case in which the cosmological dark matter is made entirely or partly made of mirror matter.
We consider scalar adiabatic primordial perturbations at linear scales in a flat Universe.
The speed of the simulations allows us for the first time to use Markov Chain Monte Carlo analyses to constrain the mirror parameters.
A Universe with pure mirror matter can fit very well the observations, equivalently to the case of an admixture with cold dark matter.
In both cases, the 
%analyses show a clear indication of 
%{\color{red}cosmological models estimate} 
cosmological models estimate 
the presence of a consistent amount of mirror dark matter, $0.06 \lesssim \Omega_{\rm mirror} h^2 \lesssim 0.12$.
\end{abstract}

\pacs{98.80.-k, 95.35.+d, 98.70.Vc, 98.65.Dx}

\keywords{cosmology, cosmic microwave background, large scale structure, dark matter}

\maketitle

%%%%%%%%%%%%%%%%%%%%%%%%%%%%%%%%%%%%%%%%%%%%%%%%%%%%%%%%%%%%%%%%%%%%%%%%%%%%%%%%

The existence of dark matter in the Universe seems to be an unavoidable evidence, as confirmed by all the currently available astrophysical observations at scales ranging from cosmological to galactic.
At the same time, its nature is still completely unknown, and is limited to its qualitative behaviour for the process of structure formation.
Together with Big Bang nucleosynthesis (BBN), the most powerful cosmological tests for dark matter candidates are the cosmic microwave background (CMB) and large scale structure (LSS) power spectra, with their increasing precisions due to the huge observational efforts of several groups.

Previous analytical and numerical studies on CMB and LSS power spectra~\cite{Ciarcelluti:2004ip,Ciarcelluti:2004ik,Ciarcelluti:2003wm,Berezhiani:2003wj,Ciarcelluti:2004ij,Ignatiev:2003js} have only limited the parameter space of mirror dark matter, without providing a definitive cosmological answer to its existence.
Here we finally address this question.

%-------------------------------------------------------------------------------

As suggested many years ago by Lee and Yang~\cite{Lee:1956qn}, to suppose the existence of mirror matter is the simplest way to restore the parity symmetry of the laws of nature.
Their idea was later developed by other authors~\cite{Blinnikov:1983gh,Kolb:1985bf,Khlopov:1989fj,Foot:1991bp}, and a lot of studies were devoted to it, showing its compatibility with all the available experimental and observational constraints (for reviews, see Refs.~\cite{Ciarcelluti:2010zz,Okun:2006eb,Ciarcelluti:2003wm,Foot:2003eq}).
In some cases, as the results of direct detection experiments~\cite{Foot:2008nw,Foot:2012rk} or the observations of neutron stars~\cite{Sandin:2008db,Ciarcelluti:2010ji}, there are interesting suggestions of the existence of mirror matter.

%-------------------------------------------------------------------------------

The original idea was to have a parallel hidden (mirror) sector of particles which is an exact duplicate of the observable sector.
In the modern context of gauge theories this implies the existence of an exact parity (mirror) symmetry between two particle sectors, that are described by the same Lagrangian and coupling constants, and consequently have the same microphysics, but where ordinary particles have left-handed interactions, mirror particles have right-handed interactions~\cite{Foot:1991bp}.
Thus, they are stable exactly as their ordinary counterparts. 

The ordinary and mirror particles have the same masses and obey to the same physical laws, but the three non-gravitational interactions act on ordinary and mirror sectors completely separately, the only link between all of them being the gravity.
Since mirror baryons do not interact with photons, or interact only very weakly, the presence of mirror matter is felt mainly by its gravitational effects, which is exactly the definition of ``dark matter''.

Hence mirror matter is a stable self-interacting\footnote{
Astrophysical constraints on self interactions of dark matter present in literature are valid only for homogeneous distributions of dark matter particles, and are therefore not directly applicable to the mirror matter case.} 
dark matter candidate that emerges if one, instead of (or in addition to) assuming a symmetry between bosons and fermions (supersymmetry), assumes that nature is parity symmetric.

Besides being a viable and powerful candidate for dark matter, the increasing interest on mirror matter is due to the fact that it provides one of the few potential explanations for the recent DAMA annual modulation signal~\cite{Foot:2008nw}, together with the results of other direct detection experiments (CDMS, CoGeNT, CRESST, XENON)~\cite{Foot:2012rk}.
The compatibility of this scenario with BBN constraints has already been studied~\cite{Ciarcelluti:2008qk,Ciarcelluti:2010dm}.

Given its consistency with experiments and observations, and the unfruitful attempts to prove the existence of the other dark matter candidates, scientific community is facing an emergent question: ``is mirror matter the dark matter of the Universe (or at least a significant part of it)?''
One possibility to answer this question is to look at the cosmological signatures of mirror particles.

It is worthwhile to note that the presence of the mirror sector does not introduce any new parameters in particle physics (if we neglect the possible weak non-gravitational interactions between visible and hidden sectors). 
But the fact that microphysics is the same in ordinary and mirror sectors does not mean that also macroscopic realizations should be the same.
The different macrophysics is usually parametrized in terms of only two ``cosmological'' free parameters: the ratio $ x $ of temperatures of the two sectors, in terms of temperatures of the ordinary and mirror photons in the cosmic background radiation; the relative amount $ \beta $ of mirror baryons compared to the ordinary ones.
\begin{eqnarray}\label{mir-param}
x \equiv \left( \frac{S'}{S} \right)^{1/3} \simeq \frac{T'}{T}
~~~~~~ {\rm and} ~~~~~~
\beta \equiv \frac{\Omega'_{\rm b}}{\Omega_{\rm b}} ~~,
\end{eqnarray}
where $T$ ($T'$), $\Omega_{\rm b}$ ($\Omega'_{\rm b}$), and $S$ ($S'$) are respectively the ordinary (mirror) photon temperature, cosmological baryon density (normalized, as usual, to the critical density of the Universe), and entropy per comoving volume~\cite{Ciarcelluti:2010zz}.

The present energy density contains relativistic (radiation) component $ \Omega_{\rm r} $, non-relativistic (matter) component $ \Omega_{\rm m} $ and the vacuum energy (cosmological term or dark energy) density $ \Omega_\Lambda $. 
According to the inflationary paradigm the Universe should be almost flat, $\Omega_{\rm tot}=\Omega_{\rm m} + \Omega_{\rm r} + \Omega_\Lambda \approx 1$, which agrees well with the results on the CMB anisotropy.
Now both radiation and matter components contain the mirror components\footnote{
Since mirror parity doubles {\it all} the ordinary particles, even if they are ``dark'' (i.e., we are not able to detect them now), whatever the form of dark matter made by some exotic ordinary particles, there will exist a mirror counterpart.
}, and the matter composition of the Universe is expressed in general by
\begin{equation}
\Omega_{\rm m} = \Omega_{\rm b}+\Omega'_{\rm b}+\Omega_{\rm DM}
         = \Omega_{\rm b} (1+\beta) + \Omega_{\rm DM}\;,
\end{equation}
where the term $\Omega_{\rm DM}$ includes the contributions of any other possible dark matter particles but mirror baryons.

At the time of BBN the mirror photons $\gamma'$, electrons ${e^{\pm}}'$ and neutrinos $\nu'_{e,\mu,\tau}$ would give a contribution to the energetic degrees of freedom equivalent to an effective number of extra neutrino families $\Delta N_\nu 
%= N_\nu -3 = (10.75/1.75) 
\simeq 6.14\,x^4$.
Current estimates of $\Delta N_\nu$~\cite{Izotov:2010ca} correspond to an upper bound $x \lesssim 0.7$, and hence at the nucleosynthesis epoch the temperature of the mirror sector should be smaller than that of the ordinary one, $T'<T$.

Due to the temperature difference between the two sectors, the cosmological key epochs take place at different redshifts, and in particular they happen in the mirror sector before than in the ordinary one~\cite{Ciarcelluti:2004ik,Ciarcelluti:2010zz}. 
The relevant epochs for the cosmic structure formation are related to the matter-radiation equality (MRE) $z_{\rm eq}$, the matter-radiation decouplings (MRD) $z_{\rm dec}$ and $z'_{\rm dec}$ due to the plasma recombinations in both sectors, and the photon-baryon equipartitions $z_{\rm b\gamma}$ and $z'_{\rm b\gamma}$. 
The MRE occurs at the redshift 
\begin{equation} \label{z-eq} 
1+z_{\rm eq}= {\frac{\Omega_{\rm m}}{\Omega_{\rm r}}} \approx 
 2.4\cdot 10^4 {\frac{\Omega_{\rm m}h^2}{1+x^4}} \;,
\end{equation}
which is always smaller than the value obtained for an ordinary Universe.
The MRD takes place in every sector only after most electrons and protons recombine into neutral hydrogen and the free electron number density diminishes, so that the interaction rate of the photons drops below the Hubble expansion rate.
Since $T'_{\rm dec} \simeq T_{\rm dec}$ up to small corrections, we obtain
\begin{equation} \label{z'_dec}
1+z'_{\rm dec} \simeq x^{-1} (1+z_{\rm dec}) 
\;, ~~
\end{equation}
so that the MRD in the mirror sector occurs earlier than in the ordinary one. 
It has been shown~\cite{Ciarcelluti:2004ik,Ciarcelluti:2010zz} that, comparing Eqs.~(\ref{z-eq}) and (\ref{z'_dec}), for $x$ smaller than a typical value $x_{\rm eq}$ the mirror photons would decouple yet during the radiation dominated period, and the evolution of primordial perturbations in the linear regime is practically identical to the standard cold dark matter (CDM) case.
Also the photon-baryon equipartition happens in the mirror sector earlier than in the ordinary one, according to the relation 
\begin{equation} \label{shiftzbg} 
1+z_{\rm b\gamma}' 
  = {\frac{\Omega_{\rm b}'}{\Omega_{\gamma}'} } 
  \simeq \frac{ \Omega_{\rm b} \, \beta}{\Omega_\gamma \, x^4} 
  = (1+z_{\rm b\gamma}) { \frac{\beta}{x^4} } > 1+z_{\rm b\gamma} \;.
\end{equation}

Previous analytical and numerical studies on CMB and LSS power spectra~\cite{Ciarcelluti:2004ip,Ciarcelluti:2004ik,Ciarcelluti:2003wm,Berezhiani:2003wj,Ciarcelluti:2004ij,Ignatiev:2003js,Foot:2012ai} have only shown, using a qualitative comparison with observations, that: (i) for low values of mirror temperatures ($x\lesssim0.3$) all the dark matter can be made of mirror baryons; (ii) for high values ($x\gtrsim0.3$) mirror baryons can be present as an admixture with CDM.

Now we are finally able to fit the cosmological parameters and obtain their quantitative estimates.

%---------------------------------------------------------

\begin{table}[h]
\caption{\label{range-par}
Adopted flat priors for the parameters.}
\begin{ruledtabular}
\begin{tabular}{ccc}
\textrm{parameter} & \textrm{lower limit} & \textrm{upper limit} \\
\colrule
$\Omega_{\rm b} h^2$     & 0.01 & 0.1\\
$\Omega_{\rm cdm} h^2$ & 0.01 & 0.8\\
$x$ & 0.05 & 0.7\\
$\beta$ & 0.5 & 9.0\\
100 $\theta_{\rm s}$ & 0.1 & 10\\
$\tau$ & 0.01 & 0.8\\
$n_{\rm s}$ & 0.7 & 1.3\\
$\ln(10^{10}A_{\rm s})$ & 2.7 & 4
\end{tabular}
\end{ruledtabular}
\end{table}

\begin{table*}[t]
\caption{\label{params}1-$\sigma$ constraints on the parameters obtained using different dark matter compositions and cosmological tests.
For the parameter $x$ we reported the upper limit computed at the 95\% c.l.}
\begin{ruledtabular}
\begin{tabular}{ccccccc}
{\sl parameter}        & standard & standard & mirror & mirror  & mirror+CDM & mirror+CDM \\
                       & CMB      & CMB+LSS  & CMB    & CMB+LSS & CMB        & CMB+LSS    \\
\hline
{\sl primary}&&&&&& \\
\hline
$\Omega_{\rm b} h^2   $&$0.02213\pm0.00041$&$0.02205\pm0.00034$&$0.02213\pm0.00040$&$0.02215\pm0.00045$&$0.02225\pm0.00044$&$0.02201\pm0.00034$\\
$\Omega_{\rm cdm} h^2 $&$0.1113 \pm0.0046 $&$0.1161 \pm0.0031 $&$ -               $&$ -               $&$0.026  \pm0.012  $&$0.036  \pm0.010  $\\
$n_{\rm s}            $&$0.9616 \pm0.0097 $&$0.9578 \pm0.0081 $&$0.966  \pm0.011  $&$0.961  \pm0.013  $&$0.964  \pm0.012  $&$0.9558 \pm0.0061 $\\
$\ln(10^{10}A_{\rm s})$&$3.051  \pm0.024  $&$3.074  \pm0.021  $&$3.082  \pm0.034  $&$3.090  \pm0.032  $&$3.100  \pm0.038  $&$3.072  \pm0.018  $\\
$100\,\theta_{\rm s}  $&$1.0406 \pm0.0017 $&$1.0404 \pm0.0015 $&$1.0413 \pm0.0016 $&$1.0403 \pm0.0016 $&$1.0408 \pm0.0017 $&$1.0400 \pm0.0014 $\\
$\tau                 $&$0.073  \pm0.012  $&$0.075  \pm0.011  $&$0.083  \pm0.014  $&$0.085  \pm0.016  $&$0.089  \pm0.016  $&$0.0755 \pm0.0085 $\\
\hline
{\sl mirror}&&&&&& \\
\hline
$x        $&$ -               $&$ -               $&$<0.456           $&$<0.297    
          $&$<0.479           $&$<0.315           $\\
%$x\; (95\%)           $&$ -               $&$ -               $&$<0.496 (0.255)   $&$<0.297 (0.227)   $&$<0.479 (0.269)   $&$<0.315 (0.242)   $\\
$\beta                $&$ -               $&$ -               $&$5.13   \pm0.30   $&$5.21   \pm0.21   $&$4.03   \pm0.50   $&$3.64   \pm0.46   $\\
\hline
{\sl derived}&&&&&& \\
\hline
$\Omega_{\rm m}       $&$0.267 \pm0.025   $&$0.292  \pm0.017  $&$0.273  \pm0.031  $&$0.285  \pm0.020  $&$0.285  \pm0.030  $&$0.291  \pm0.017  $\\
$\Omega_\Lambda       $&$0.733 \pm0.025   $&$0.708  \pm0.017  $&$0.727  \pm0.031  $&$0.715  \pm0.020  $&$0.715  \pm0.030  $&$0.709  \pm0.017  $\\
$z_{\rm re}           $&$9.3   \pm1.0     $&$9.70   \pm0.95   $&$10.3   \pm1.2    $&$10.5   \pm1.3    $&$10.8   \pm1.4    $&$9.73   \pm0.77   $\\
$h                    $&$0.710 \pm0.022   $&$0.690  \pm0.013  $&$0.708  \pm0.024  $&$0.695  \pm0.017  $&$0.698  \pm0.023  $&$0.690  \pm0.014  $\\
age [Gyr]              &$13.750\pm0.092   $&$13.793 \pm0.066  $&$13.687 \pm0.093  $&$13.759 \pm0.088  $&$13.71  \pm0.10   $&$13.782 \pm0.067  $\\
$\sigma_8             $&$ -               $&$0.824  \pm0.015  $&$ -               $&$0.767  \pm0.021  $&$ -               $&$0.746  \pm0.018  $\\
\end{tabular}
\end{ruledtabular}
\end{table*}

We have modified the publicly available cosmological simulation tools CAMB~\cite{Lewis:1999bs} and CosmoMC~\cite{Lewis:2002ah} in order to include the effects of mirror matter.
Since the physics of the mirror particles is the same as our particles, we have doubled the equations separately in each sector, and considered all the particles when describing the gravitational interactions.
The recombinations are computed separately for each sector.
The computational times of this modified version of CAMB are considerably increased, but still fast enough to compute the many models needed for a Monte Carlo fit in reasonable times.

Compared with previous numerical simulations~\cite{Ciarcelluti:2004ip,Ciarcelluti:2004ik,Ciarcelluti:2003wm,Berezhiani:2003wj,Ciarcelluti:2004ij}, we have used an updated estimate of the primordial chemical composition of mirror particles present in Refs.~\cite{Ciarcelluti:2008vm,Ciarcelluti:2008vs,Ciarcelluti:2009da,Ciarcelluti:2010zz,Ciarcelluti:2014vta}, and a more accurate treatment of the recombinations of ordinary and mirror particles using the numerical code RECFAST~\cite{Seager:1999bc}.
The new models based on CAMB and the more accurate treatment of mirror BBN are consistent with the previous ones, but there is a strong improvement of the computational time, allowing us now to constrain the parameters.

We use a Markov Chain Monte Carlo (MCMC) sampling of the multi-dimensional likelihood as a function of model parameters, based on the computations of CMB and LSS power spectra obtained with our modified version of CAMB.

We sample the following eight-dimensional set of cosmological parameters,
adopting flat priors on them 
with broad distributions, 
as shown in Table~\ref{range-par}: 
the baryon and cold dark matter densities $\Omega_{\rm b} h^2$ 
and $\Omega_{\rm cdm} h^2$, 
the relative mirror photon temperature $x$, 
the relative mirror baryon density $\beta$, 
the ratio of the sound horizon to the angular diameter distance at decoupling $\theta_{\rm s}$, 
the reionization optical depth $\tau$, 
the scalar spectral index $n_s$ and 
the scalar fluctuation amplitude $A_s$.
The upper limit on $x$ is set by the aforementioned BBN limit.
In addition, we obtain constraints on derived parameters: 
the matter and dark energy densities normalized to the critical density $\Omega_{\rm m}$ 
and $\Omega_{\lambda}$, 
the reionization redshift $z_{\rm re}$, 
the Hubble parameter $h$, 
the age of the Universe in Gyr, 
the density fluctuation amplitude $\sigma_8$ at $8 h^{-1} {\rm Mpc}$.
The runs also include weak priors on the Hubble parameter, $0.4 \le h\le 1.0$, 
and on the age of the universe, $10 \le {\rm age (Gyr)} \le 20$.
In all the computations we assume scalar adiabatic initial conditions in a flat Universe ($\Omega_{\rm tot}=1$), a dark energy equation of state with $w=-1$, massless neutrinos, and the number of neutrino families of the standard model $N_{\rm eff}=3.046$.

\begin{figure}[t]
\includegraphics[scale=0.52]{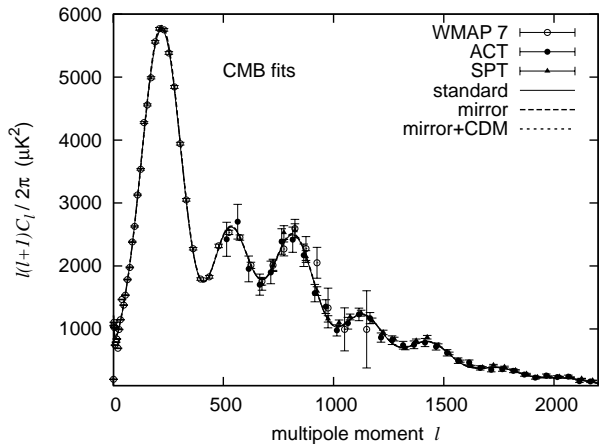}
\caption{\label{fits_CMB}  CMB power spectrum for best-fit models with baryons and mirror matter (dashed line), or baryons, mirror matter, and cold dark matter (dotted line) obtained using CMB only data.
For comparison we show also the standard model fit (solid line).}
\end{figure}

\begin{figure}[t]
\includegraphics[scale=0.52]{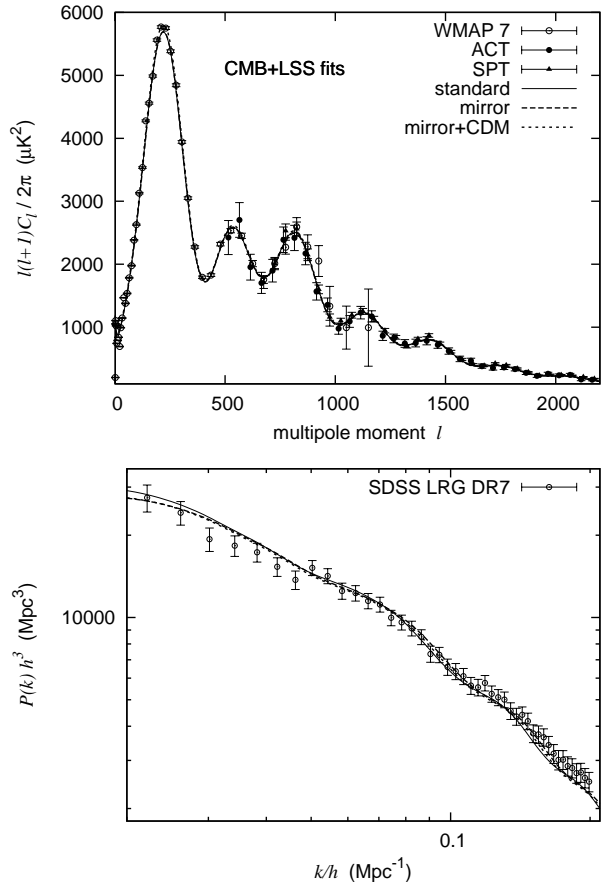}
\caption{\label{fits_CMBLSS} CMB and LSS power spectra for best-fit models with baryons and mirror matter (dashed line), or baryons, mirror matter, and cold dark matter (dotted line) obtained using both CMB and LSS datasets.
For comparison we show also the standard model fit (solid line).}
\end{figure}

We consider two different chemical compositions of dark matter: the case pure mirror and the case mixed mirror-CDM.
In addition, we perform analyses using two different configurations: the CMB only and the CMB combined with the LSS.
The CMB datasets are provided by the WMAP7 team~\cite{Larson:2010gs}, which measured the acoustic oscillations of the primordial plasma on degree scales with cosmic-variance-limited precision, together with the ACT~\cite{Sievers:2013wk} and SPT~\cite{Keisler:2011aw} observations, which provided accurate power spectra at higher {\sl l}'s.
For the LSS, instead, we include the power spectrum extracted from the SDSS-DR7 luminous red galaxy sample~\cite{Reid:2009xm} limited to the length scales larger than $k \sim 0.2 h\:{\rm Mpc^{-1}}$ to avoid non-linear clustering and scale-dependent galaxy biasing effects.
For comparison, we run also a MCMC chain for a standard $\Lambda$CDM cosmology, with the same assumptions and priors, to use as a reference model.
The results of the runs are shown in Table~\ref{params}, where the estimates of the parameters and the 1-$\sigma$ confidence intervals are obtained by marginalizing the multi-dimensional likelihoods down to one dimension.
For the parameter $x$, it is just possible to obtain an upper limit, since the probability density of that parameter is almost flat in the low-$x$ region, while it sharply decreases at higher $x$. 
We choose to give the upper limits at the $95\%$ confidence level.
%for $x$ smaller than a value, for which there is a small statistical preference, while for higher values it goes towards zero.
This is compatible with what theoretically expected, since, as previously studied~\cite{Ciarcelluti:2004ik,Ciarcelluti:2004ip,Ciarcelluti:2010zz}, for smaller $x$ the decoupling of mirror baryons happens at earlier times, mimicking more and more the usual CDM behaviour at linear regimes.
The corresponding best-fit models are shown in Figs.~\ref{fits_CMB} and \ref{fits_CMBLSS} for analysis based respectively on CMB only and on both CMB and LSS.

Looking at Table~\ref{params}, we see that the values of the primary cosmological parameters, except obviously for the CDM density, do not vary significantly between the standard model and both the pure and mixed mirror compositions, for both kinds of analysis (CMB and CMB+LSS).
Going to the derived parameters, there is an increase of the matter content of the Universe at the expenses of the dark energy for models obtained considering the CMB only.
This is partly due to a bigger matter density, and partly to the decrease of the Hubble parameter, that is always compatible with the current estimates.
For all the models, the non-baryonic matter density is between 5 and 6 times the baryonic density, in accordance with common cosmological analyses.
But the most interesting result is concentrated in the lines constraining the mirror parameters.
Concerning $x$, the CMB only analysis estimates 
at $95\%$ c.l. an upper bound $x<0.456$
%a central value 
for a pure mirror model,
% $x\simeq0.25$, 
while for mixed mirror this 
bound becomes slightly weaker, $x<0.479$.
%value is slightly bigger, $x\simeq0.27$. 
Both these 
allowed regions
%values lie in the parameter region
include the values 
able to explain the results of the dark matter direct detection experiments~\cite{Foot:2008nw,Foot:2012rk}.
The inclusion of the LSS 
significantly tightens the allowed region 
%slightly lowers the estimates 
of $x$ for both pure 
($x<0.297$)
and mixed 
($x<0.315$)
mirror, confirming the higher sensitivity of the LSS on $x$ already evidenced in previous works~\cite{Ciarcelluti:2004ip,Ciarcelluti:2010zz}.
Even these more stringent constraints are compatible with the mirror matter interpratation of direct detection experiments~\cite{Foot:2008nw,Foot:2012rk}.
We finally look at the most significant parameter, $\beta$, which expresses how much, if any, mirror matter is present in the Universe.
The results obtained using CMB alone or combined with the LSS are similar.
For pure mirror this value is between 5 and 5.5, 
%clearly showing that cosmology requires a strong presence 
%{\color{red}showing that mirror cosmological models require the presence of a large amount} 
showing that mirror cosmological models require the presence of a large amount 
of mirror matter in order to interpret its observables.
In case of mirror mixed with CDM, the results show densities of mirror matter that are between 2 and 4 times larger than those of CDM.
This is an 
%important 
%{\color{red}interesting}
interesting
result, 
%since it means that, when it can freely choose between mirror dark matter or collisionless massive WIMPs, cosmology unequivocally requires the presence of a dominant fraction of mirror matter.
%{\color{red}suggesting that mirror matter could contribute significantly to the matter budget of the Universe in a similar way as CDM does.
%Future data, especially on LSS, should help to discriminate between mirror and CDM models.}
suggesting that mirror matter could contribute significantly to the matter budget of the Universe in a similar way as CDM does.
Future data, especially on LSS, should help to discriminate between mirror and CDM models.

The likelihoods of the best fit models have very similar values for each class of models obtained using CMB only or the combination of CMB and LSS data.
In the former case, they are $-\ln({\cal L})=3772$ for pure CDM, $-\ln({\cal L})=3771$ for pure mirror and $-\ln({\cal L})=3771$ for the mixture mirror-CDM, while for the latter they are respectively 3795, 3794 and 3795.
Considering the increases of one or two free parameters between the models, these values don't show statistically significant differences.

In Figs.~\ref{fits_CMB} and \ref{fits_CMBLSS} the agreement of the best fits with the data is shown, together with the comparison with the reference standard model.
For the CMB the models obtained in both the analyses, namely fitting CMB data alone or CMB combined with LSS, are almost indistinguishable between themselves, and few differences are present in the LSS power spectra.

To summarize, in this work we have obtained two main results.
First of all, for the first time the two parameters describing the mirror matter are constrained.
Considering the most stringent analyses performed using both the CMB and LSS, we obtained $x<0.297$ ($95\%$ c.l.) and $\beta=5.21\pm0.21$ (1$\sigma$) for pure mirror, $x<0.315$ ($95\%$ c.l.) and $\beta=3.64\pm0.46$ (1$\sigma$) for mixed mirror-CDM.
These 
bounds include 
%values lie in 
the range 
of parameters 
required for interesting consequences on observations and experiments.
Secondly, 
%but even more important, the analyses show a clear indication of the presence of consistent amounts of mirror dark matter in the Universe.
%{\color{red}we have demonstrated that cosmological models with pure mirror matter, mirror matter mixed with CDM and pure CDM are equivalent concerning the CMB and LSS power spectra, as a consequence of the fact that mirror matter and collisional WIMPs have the same behaviour at linear scales.}
we have demonstrated that cosmological models with pure mirror matter, mirror matter mixed with CDM and pure CDM are equivalent concerning the CMB and LSS power spectra, as a consequence of the fact that mirror matter and collisional WIMPs have the same behaviour at linear scales.

\begin{acknowledgments}
We are grateful to Jean-R\'en\'e Cudell for useful discussions, and to an anonymous reviewer for improving the statistical analysis.
PC acknowledges the hospitality of the IFPA group and the financial support of the Belgian Science Policy during part of this work.
QW is supported by the Belgian Fund FRS-FNRS as a Research Fellow.
\end{acknowledgments}

%%%%%%%%%%%%%%%%%%%%%%%%%%%%%%%%%%%%%%%%%%%%%%%%%%%%%%%%%%%%%%%%%%%%%%%%%%%%%%%%

\bibliography{fit-cosmo-mir-v5}% Produces the bibliography via BibTeX.

\end{document}